\documentclass{PoS}
\usepackage{amsmath}
\usepackage{lipsum}
\usepackage{enumitem}
\usepackage{soul}

\newcommand{\TeV}{{\ensuremath\rm TeV}}
\newcommand{\GeV}{{\ensuremath\rm GeV}}
\newcommand{\MeV}{{\ensuremath\rm MeV}}

\newcommand{\pb}{{\ensuremath\rm pb}}
\newcommand{\eqn}{equation}

\newcommand{\lam}{\lambda}

\def\D0{\slash\!\!\!\!\!\!\!\!\!\:D0}

\newcommand{\oblique}{Altarelli:1990zd,Peskin:1990zt,Peskin:1991sw,Maksymyk:1993zm}

\title{Inert Doublet Model in the light of LHC and astrophysical data }

\ShortTitle{Inert Doublet Model in the light of LHC and astrophysical data }

\author{\speaker{Agnieszka Ilnicka}$^{a,b}$,Maria Krawczyk$^c$, Tania Robens$^d$\\
$^a$ Institute of Physics, University of Z\"{u}rich \\
Winterthurstrasse 190, CH-8057 Z\"{u}rich\\
$^b$ Physics Department, ETH Z\"{u}rich \\
Otto-Stern-Weg 5, CH - 8093 Z\"{u}rich\\
$^c$ Faculty of Physics, University of Warsaw \\
 ul. Pasteura 5, 02-093 Warsaw\\
 $^d$ TU Dresden, Institut f\"{u}r Kern- und Teilchenphysik\\
  Zellescher Weg 19, D-01069 Dresden \\
E-mail: \email{ailnicka@physik.uzh.ch},\email{maria.krawczyk@fuw.edu.pl},\email{Tania.Robens@tu-dresden.de }}

\abstract{
{We}  investigate the parameter space of the Inert Doublet Model, which {is} a straightforward extension of the SM in the {scalar} sector. We apply a set of{ constraints} both from the theoretical and experimental side to extract and determine allowed regions of parameter space. These  constraints put strong limits on {both} masses and couplings of the new particles. We also present {a} set of benchmarks {for the current LHC run}. This work is based on \cite{Ilnicka:2015sra,Ilnicka:2015jba}.
}

\FullConference{The European Physical Society Conference on High Energy Physics\\
		22--29 July 2015\\
		Vienna, Austria}

\begin{document}

\section{Introduction}

The Inert Doublet Model (IDM) is one of the {most straightforward} extensions of the Standard Model. In this model,  the scalar sector is augmented by a second complex doublet, and an exact $Z_2$ symmetry is imposed on the Lagrangian. The first doublet corresponds to the SM Higgs{ doublet}  with the Higgs particle, and is responsible for electroweak symmetry breaking (EWSB) in a standard way. The second doublet,{ called dark doublet,}  contains a stable dark matter candidate. {The} IDM was studied in context of LHC {phenomenology}, both {with respect to the} Higgs boson discovery \cite{Barbieri:2006dq,Cao:2007rm} as well as dark matter discovery, {the latter} e.g. in {the} two lepton + MET channel \cite{Dolle:2009ft,Gustafsson:2012aj}.  Moreover the model offers also rich cosmological phenomenology,
for {a} review of references see \cite{Ilnicka:2015jba}.

The discovery of a Higgs boson in 2012 basically fixes the particle content of the first{ doublet}
in the IDM, in analogy to the scalar sector of the SM. \footnote{{After its discovery, several analyses studied the impact of this discovery as well as the signal strength measurements on the particles parameter space  \cite{Swiezewska:2012eh,Gustafsson:2012aj,Arhrib:2013ela,Krawczyk:2013jta}}. } In the light of LHC run II knowledge about regions of parameter space which are in agreement with all current constraints is imminent in order to correctly determine open search channels and their experimental signatures. We address this need by presenting a {complete} survey on the model's parameter space including a wide range of constraints, {coming} from theoretical bounds as well as collider and astrophysical {data}.  We additionally provide a set of benchmark points and planes for the current LHC run.

\section{The Model}

The scalar sector of IDM consists of two doublets of complex scalar fields, which we label $\phi_S$ {for the SM-like doublet}, and $\phi_D$ - {for the dark doublet}. After EWSB, only $\phi_S$ acquires a nonzero vacuum expectation value ($v$). 
The $Z_2$ - symmetric potential reads:
\begin{equation}\begin{array}{c}
V=-\frac{1}{2}\left[m_{11}^2(\phi_S^\dagger\phi_S)\!+\! m_{22}^2(\phi_D^\dagger\phi_D)\right]+
\frac{\lambda_1}{2}(\phi_S^\dagger\phi_S)^2\! 
+\!\frac{\lambda_2}{2}(\phi_D^\dagger\phi_D)^2\\[2mm]+\!\lambda_3(\phi_S^\dagger\phi_S)(\phi_D^\dagger\phi_D)\!
\!+\!\lambda_4(\phi_S^\dagger\phi_D)(\phi_D^\dagger\phi_S) +\frac{\lambda_5}{2}\left[(\phi_S^\dagger\phi_D)^2\!
+\!(\phi_D^\dagger\phi_S)^2\right],
\end{array}\label{pot}\end{equation}
with {the} $Z_2$ transformation {being defined by} $\phi_S \rightarrow \phi_S, \phi_D \rightarrow - \phi_D, SM \rightarrow SM$. Due to this symmetry 
{the} lightest particle of {the} dark sector is stable. {In total}, the dark sector {contains} 4 new particles: $H$, $A$ and $H^{\pm}$. We here choose H to  be  the dark matter (DM) candidate\footnote{A priori, any of the new scalars can function as a dark matter candidate. However, we neglect the choice of a charged dark matter candidate, as these are strongly constrained \cite{Chuzhoy:2008zy}. Choosing A instead of H changes the meaning of $\lam_5$, but not the overall phenomenology of the model, cf. \cite{Ilnicka:2015jba}.}.  \\
The Higgs boson {data}  and electroweak precision observables fixes the SM-like Higgs mass $M_h$ and $v$, {and we are left with} 5 free parameters, {for which we take}: 
\begin{\eqn}\label{eq:physbas}
M_H, M_A, M_{H^{\pm}}, \lam_2, \lam_{345},
\end{\eqn}
where { the $\lam_{345} = \lam_{3}+\lam_{4}+\lam_{5}$ describes { coupling between SM-like Higgs and DM particle H}
and $\lambda_2$ corresponds to self-couplings of the dark scalars}.

\section{Scan Procedure}
\subsection{Constraints}
Here we enumerate the constraints which were included in our analysis 
 \cite{Ilnicka:2015jba}.\\
\paragraph{Theoretical constraints:}
\begin{itemize}
\item{} positivity: potential is bounded from below at tree level
\item{} perturbative  unitarity of 2 $\to$ 2 scalar scattering matrix
\item{} perturbativity of all couplings, {(we chose $4\,\pi$ as upper limit)}
\item{} condition to be in the inert vacuum \cite{Swiezewska:2012ej,Gustafsson:2010zz}.
\end{itemize}
\paragraph{Experimental constraints}
\begin{itemize}
\item{} Mass of the SM-like Higgs boson $h$ set to $M_h=125.1\,\GeV $ in agreement with results from the LHC experiments \cite{Aad:2015zhl} 
\item{} Total width of the $h$ obey an upper limit $\Gamma_\text{tot}\,\leq\,22\,\MeV$ \cite{Khachatryan:2014iha,Aad:2015xua}
\item{} Bounds {provided by} the total width of the electroweak gauge bosons:
$$M_{A,H}+M_H^\pm\,\geq\,m_W,\,M_A+M_H\,\geq\,m_Z,\,2\,M_H^\pm\,\geq\,m_Z$$
\item{} Bound on the lower mass of $M_H^\pm\,\geq\,70\,\GeV$  \cite{Pierce:2007ut}.
\item{}Agreement with the current null-searches from the LEP, Tevatron, and LHC experiments {using} HiggsBounds \cite{Bechtle:2008jh, Bechtle:2011sb, Bechtle:2013wla}
\item{} Agreement within $2\,\sigma$ for the 125 GeV~ Higgs signal strength measurements {using} HiggsSignals \cite{Bechtle:2013xfa}
\item{} $2\,\sigma$ agreement with electroweak precision observables, parameterized through the (correlated) electroweak oblique parameters $S,T,U$ \cite{\oblique}.
\item{} Upper limit of { the $H^+$ }lifetime $\tau\,\leq\,10^{-7}\,s$, leading to $\Gamma_{\text{tot}}\,\geq\,6.58\,\times\,10^{-18}\,\GeV$.
\item{} Upper limit on relic density within $2\sigma$ from measurement of the Planck experiment \cite{Planck:2015xua}: $\Omega_c\,h^2\,\leq\, 0.1241$
\item{} Respect direct detection limits from dark matter nucleon scattering: the most stringent bounds are  provided by the LUX experiment \cite{Akerib:2013tjd}
\item{} Obey exclusions from recasted SUSY LEP and LHC {analyses} \cite{Lundstrom:2008ai,Belanger:2015kga} 
\end{itemize}

\subsection{Scan setup}
We have performed a {flat} scan over the input parameters as given in eq. (\ref{eq:physbas}), where we have run up {1 \TeV~ with the mass of the DM candidate}. We confronted these points with the above constraints{ (sec. 3.1)}, where for each point we memorised the exclusion criteria if applicable. Our scan is performed in several steps: first we check the theoretical constraints as well as $S,T,U$ parameters and total widths using 2HDMC \cite{Eriksson:2009ws}. In a second step, points which pass these bounds were confronted with null searches and  Higgs signal strength measurements using HiggsBounds and HiggsSignals, respectively \cite{Bechtle:2008jh, Bechtle:2011sb, Bechtle:2013wla,Bechtle:2013xfa}. In the final step of the scan, the calculation of Dark Matter observables, {i.e.} the total relic density as well as the direct detection cross section, was performed using MicrOmegas \cite{Belanger:2013oya} and confronted with Planck and LUX measurements \cite{Planck:2015xua,Akerib:2013tjd}. For points which {are in agreement with all bounds}, we provide cross sections for pair-production {of dark scalars}, {where we employed}  MadGraph \cite{Alwall:2011uj}, {using the IDM UFO model presented in \cite{Goudelis:2013uca}}.

\section{Allowed Parameter Space of IDM}

\begin{figure}[h]
\begin{center}
\includegraphics[width=0.48\textwidth]{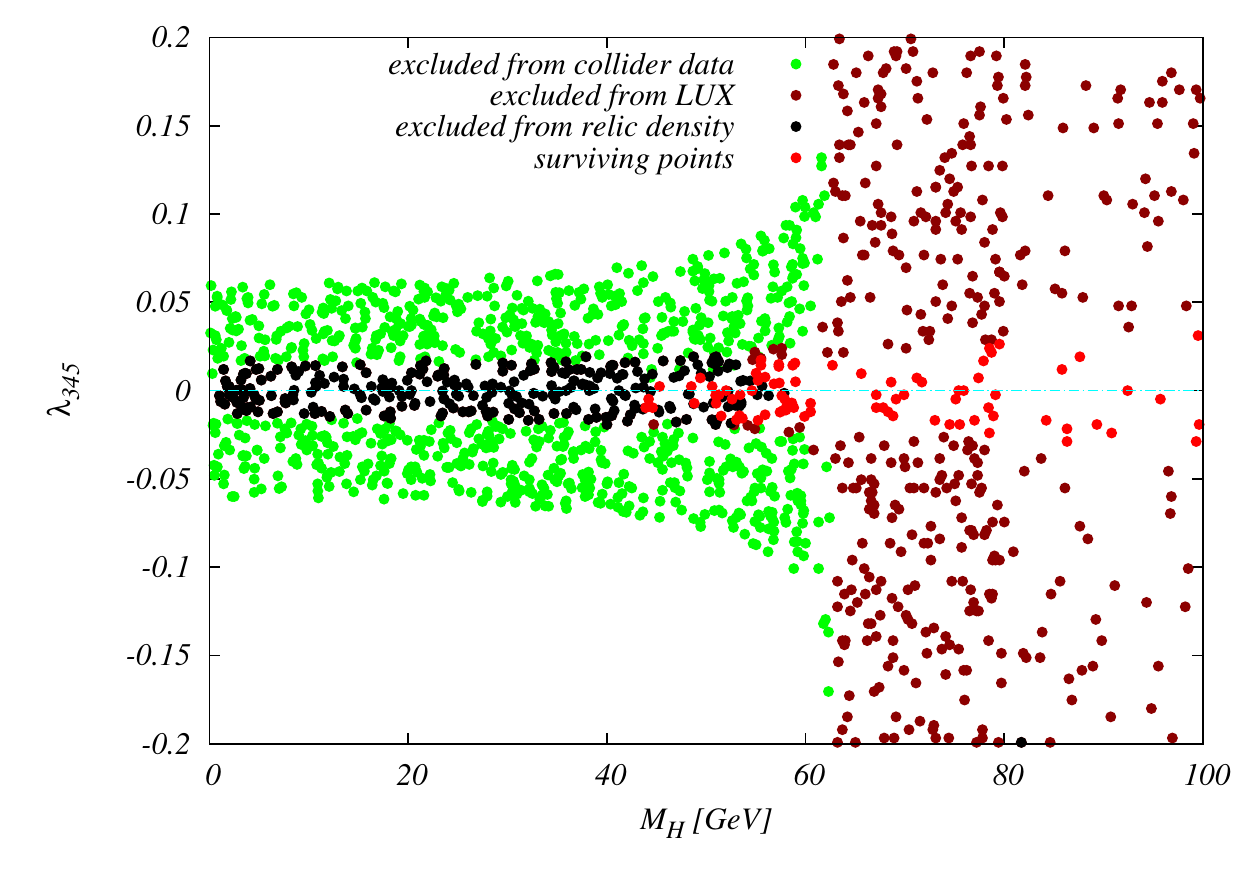}
\includegraphics[width=0.48\textwidth]{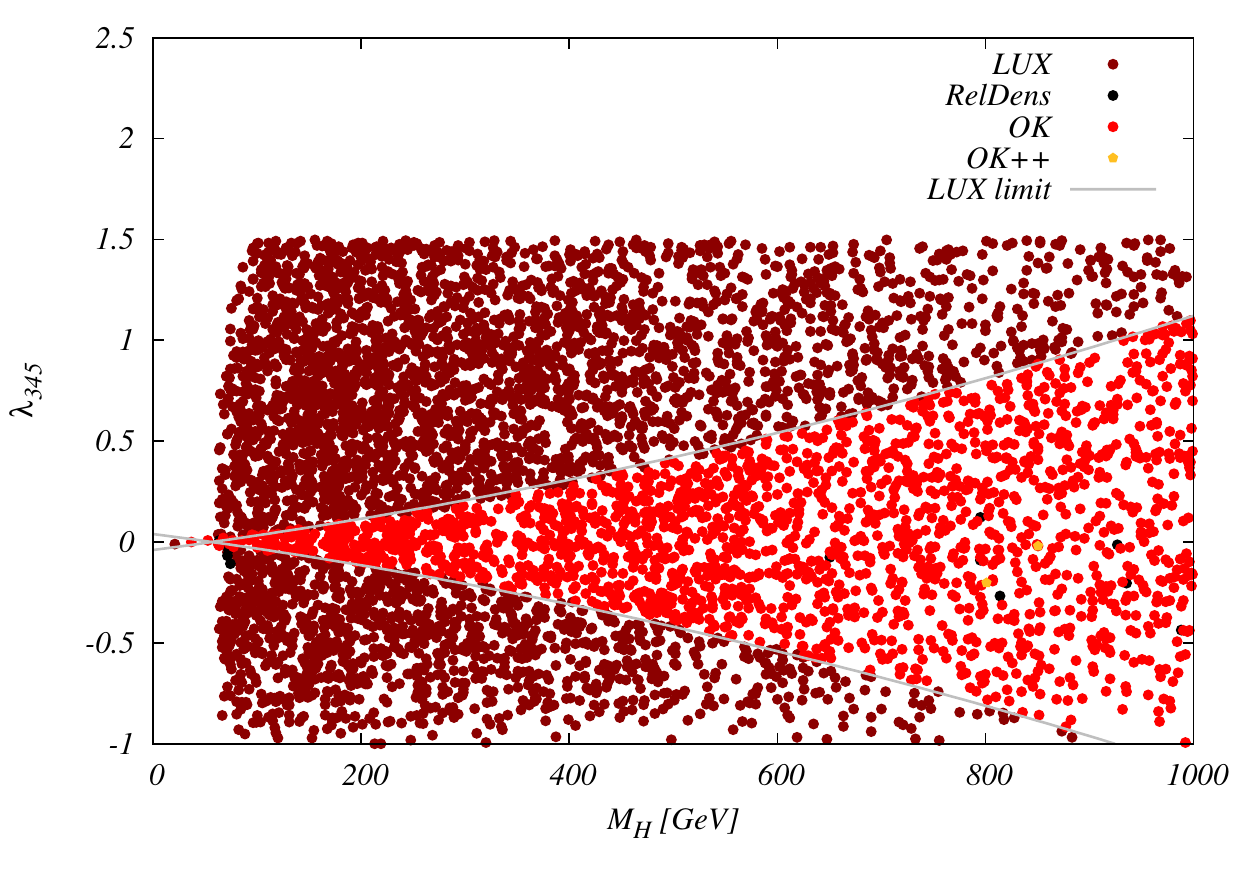}
\caption{\label{fig:lam345} Dark matter mass vs DM-SM coupling planes: {\sl (left)} {a zoom into the}  low mass region, {with $M_H\,\leq\,100\,\GeV$} showing {the importance of} constraints from LHC and astrophysical measurements; {\sl (right)} {the} general case { with $M_H$ up tp 1 TeV}, where bounds from direct detection are {dominant}.}
\end{center}
\end{figure}

{In this section,}  we present the results of our scans {and} we emphasise the source of the strongest bounds, {following the order of checks as discussed above}. In {the} left panel {of} Figure \ref{fig:lam345} we  {show} the  region of parameter space {where the dark scalar mass $M_H$ is smaller than the SM like Higgs mass of 125 \GeV.} {This region} is strongly constrained by the combination of {the 125 \GeV~Higgs} width {and the signal strength}, leaving {a} very narrow allowed {stripe of} $\lam_{345}$, {with absolute values $\lesssim\,0.02$}. Also astrophysical data {pose} important {constraints} in this region. Relic density requires {the mass of the DM candidate} to values $\gtrsim\,45\,\GeV$. In addition, LUX measurements narrow down {the} allowed values for the $\lam_{345}$ coupling. {In the right panel of Figure \ref{fig:lam345}, we display the whole region of mass values for $M_H$ up to 1 \TeV.} {Here, especially}  LUX{ data} {limit the models parameter space}, reflecting the dependance {of the direct detection cross section on} {these parameters}.

\begin{figure}[h]
\begin{center}
\includegraphics[width=0.48\textwidth]{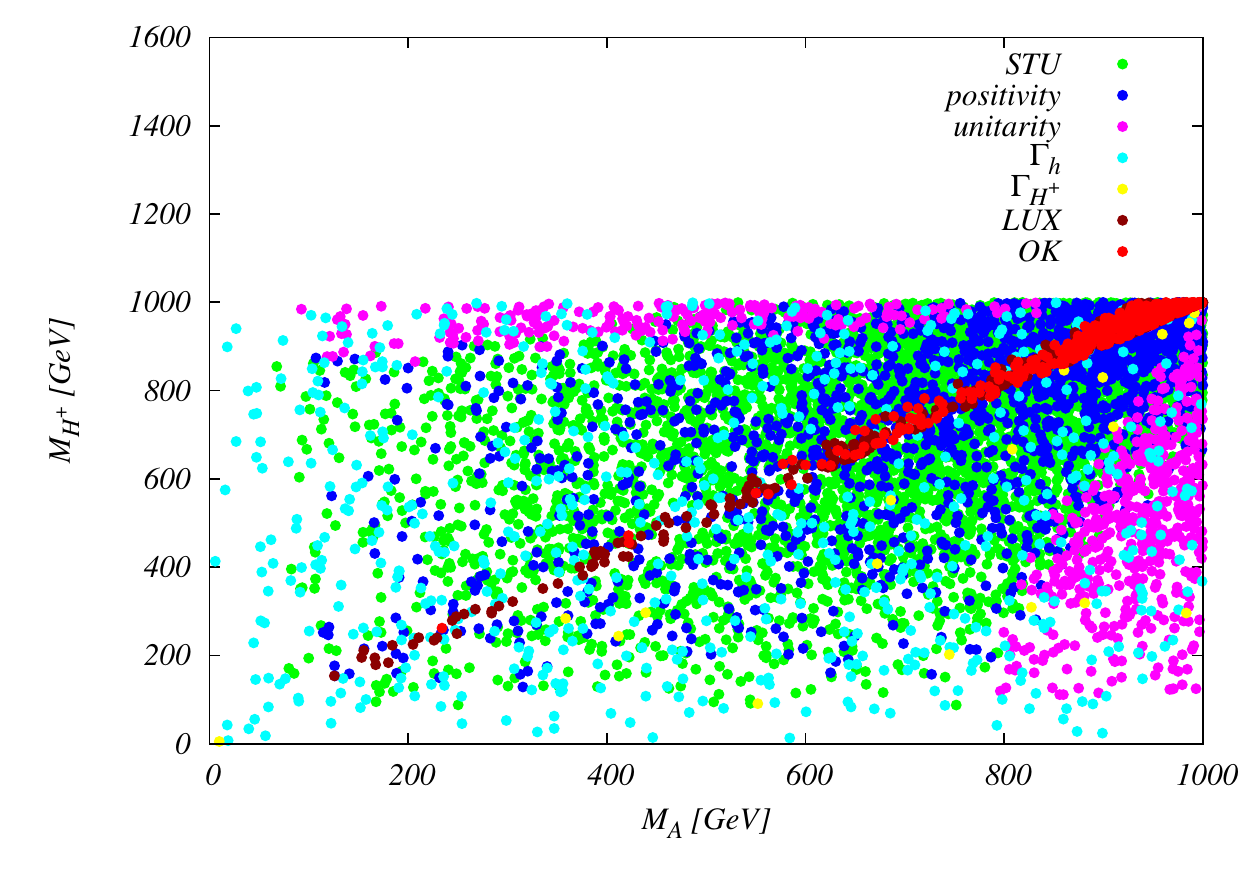}
\includegraphics[width=0.48\textwidth]{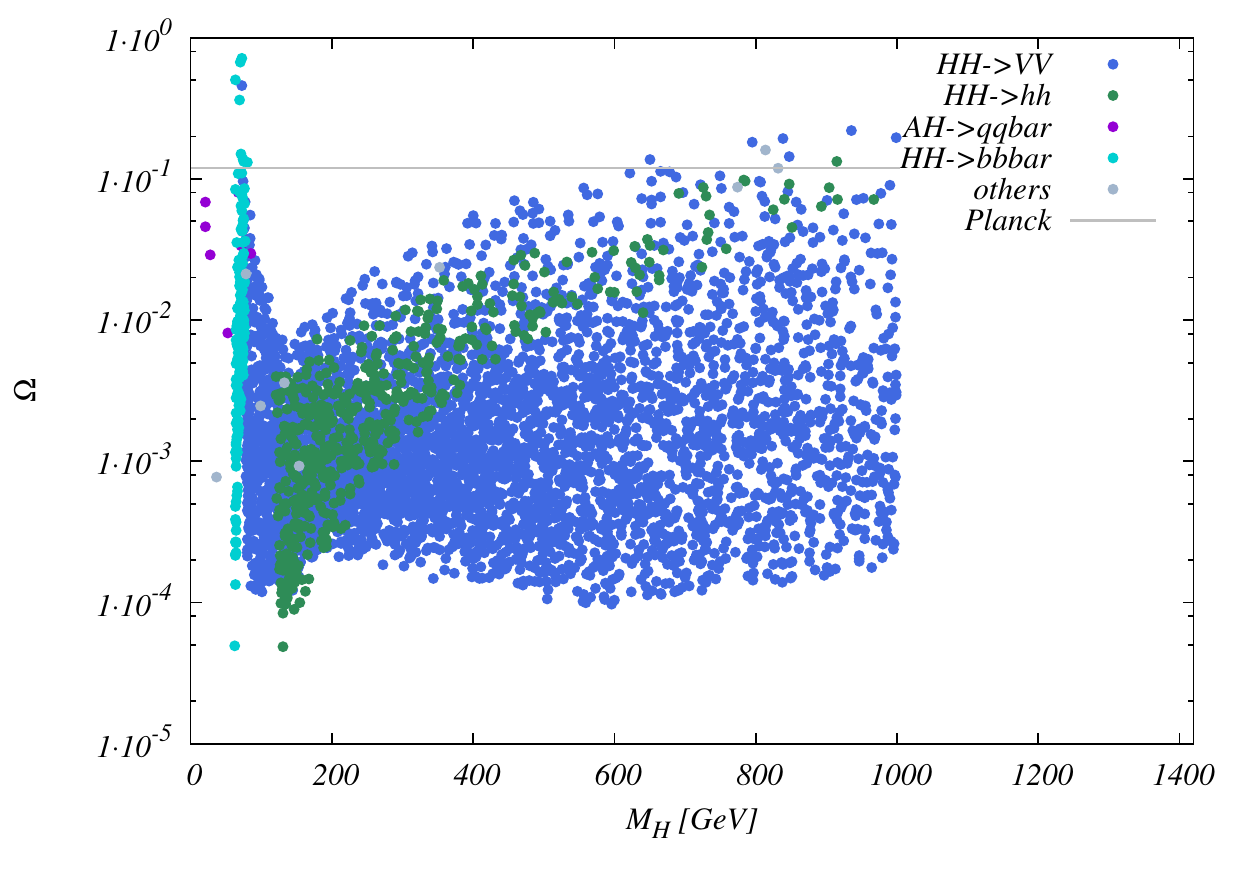}
\caption{\label{fig:other} {\sl (left)} The exclusion plot in $M_A$ vs $M_{H^+}$ plane. {\sl (right)} The plot presents leading contribution to DM relic density in the DM mass vs relic density plane. (Exclusion from LUX not included).}
\end{center}
\end{figure}

The interplay of {all} constraints, presented in {the} left panel of Figure \ref{fig:other}, leads to the strict mass hierarchy
$$M_H < M_A \leq M_{H^+}.$$
Also clearly visible is {the} preference {for} more degenerate masses in the dark sector and the more relaxed parameter space for high masses.  

In right panel of Figure \ref{fig:other} we present the leading contributions to the DM relic density. The relic density produced by IDM particles can be large for very low or relatively high DM masses. {The} dominant channel for most of the parameter space{ for mass of H above 100 GeV}  is annihilation into vector bosons.

\section{Benchmarks}

\begin{figure}[h]
\begin{center}
\includegraphics[angle=-90,width=0.48\textwidth]{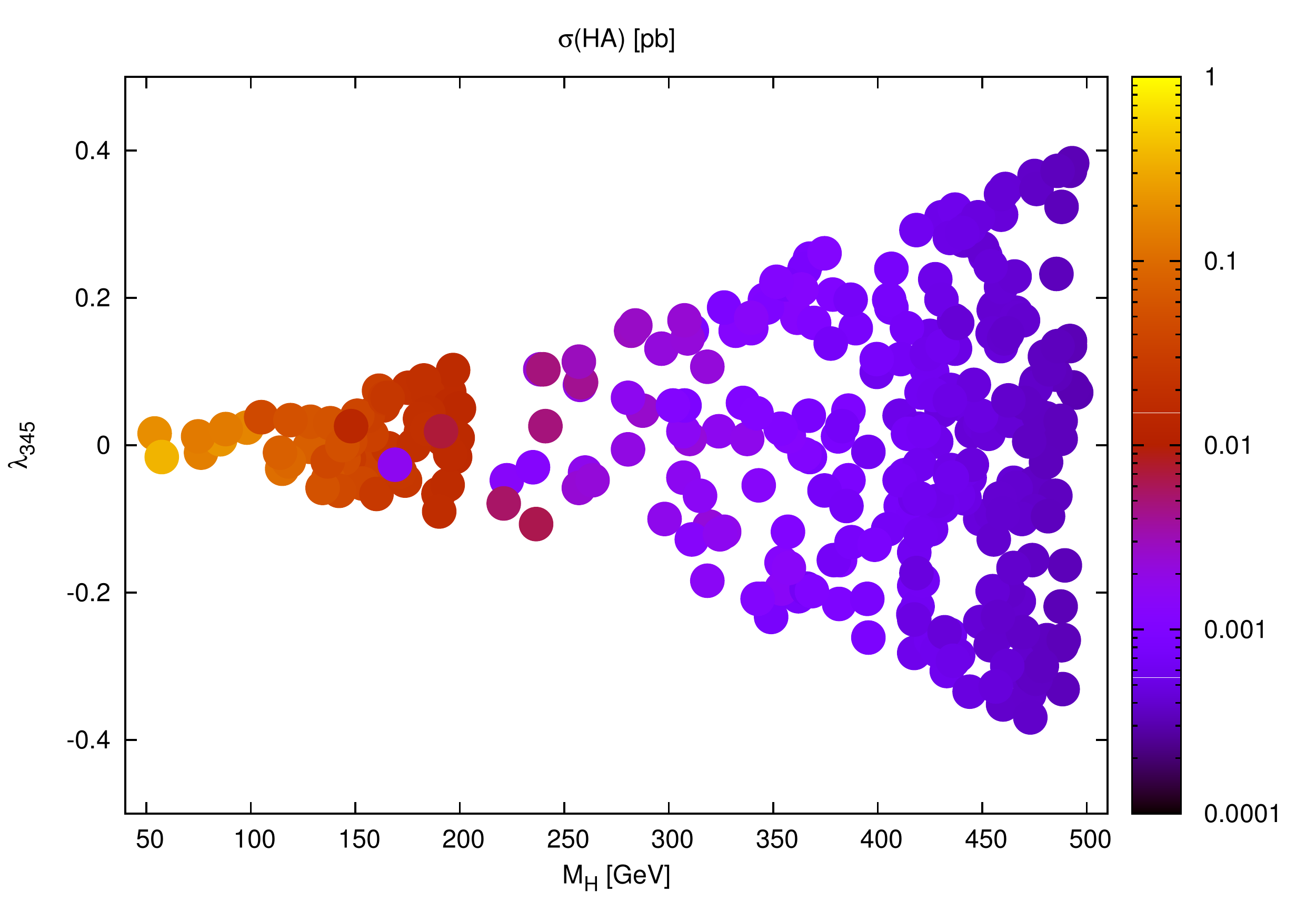}
\includegraphics[angle=-90,width=0.48\textwidth]{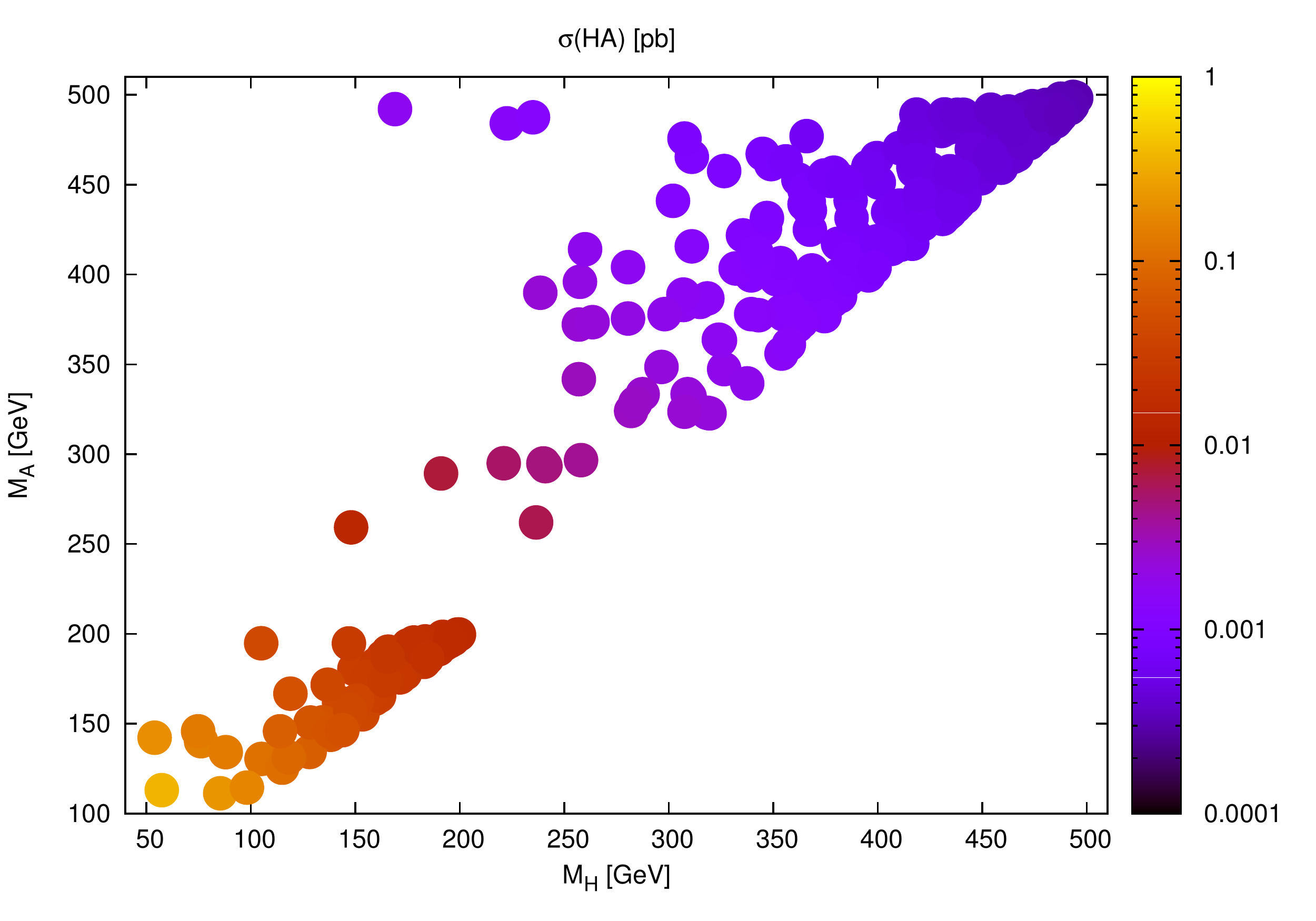}
\caption{\label{fig:plane} The planes of allowed points, {with{ HA} production cross sections (in  \pb) at a 13 \TeV~LHC.}}
\end{center}
\end{figure}

For the points allowed by all constraints the leading order cross section { dark scalars }pair-production \footnote{Due to the $Z_2$ symmetry dark sector particles are always produced in pairs.} was calculated. The dominating channel {leading to visible collider signatures} is HA production \footnote{As most DM models at colliders, the IDM will always lead to signatures including missing transverse energy.}, and Figure \ref{fig:plane} {shows} its dependence on {the} masses and coupling of dark matter. {Figure} \ref{fig:plane} {shows} that the production cross section is {mainly} driven by {kinematics, and especially by} particles masses. {This follows from the fact that t}he dominant production mechanism, namely production via Z mediation in s-channel, only depends on kinematics and the SM electroweak couplings, {but not on couplings of the extended scalar sector which are absent in the SM}.

From the presented points, five were chosen as benchmarks \cite{Ilnicka:2015sra,usbm,Ilnicka:2015jba}, see Table \ref{tab:bmk}. While benchmarks I and II are exceptional points in a sense that the allowed parameter space is extremely constrained in the low mass region, benchmarks III to V are more typical, as these parts of the parameter space are more highly populated. Furthermore, for scenario IV the production cross sections for $HA$ and $H^+ H^-$ have similar order of magnitude.

\begin{table}[h]
\small
\begin{tabular}{l c c c c c}
\hline \hline 
BP & \bf{BP I} & \bf{BP II} & \bf{BP III} & \bf{BP IV} & \bf{BP V} \\
\hline \hline
$M_H [GeV]$ & 57.5 &85.5 & 128.0 & 363.0 & 311.0 \\
$M_A [GeV]$ & 113.0 & 111.0 &  134.0 & 374.0 &  415.0 \\
$ M_{H^\pm} [GeV]$ & 123.0 & 140.0  & 176.0 & 374.0 & 447.0 \\
$|\lam_{345}|$ &  [0.002;0.015] & [0;0.015] &  [0;0.05] & [0,0.25] &[0;0.19] \\
$|\lam_2|$ & [0,4.2] &[0,4.2] &[0,4.2] &[0,4.2] &[0,4.2] \\
$\sigma(pp \rightarrow HA) [pb]$ & 0.371(4) & 0.226 (2) &0.0765 (7) & 0.00122(1) & 0.00129 (1)\\
$\sigma(pp \rightarrow H^+H^-) [pb]$& 0.097 (1) &  0.0605 (9) &0.0259 (3) &0.00124 (1)  & 0.000553 (7) \\
$BR(H^+\rightarrow H W^+$ & 0.99 & 0.96 &  0.66 & 1 &   0.99 \\
\hline
\end{tabular}
\label{tab:bmk}
\caption{Benchmark points for{ dark scalars pair production at }  LHC run 2. }
\end{table}

\section{Conclusions}
{The} Inert Doublet Model is {a} very promising extension of {the} SM {in the scalar sector}. {Its parameter space} is subject to {several theoretical and experimental} constraints. {In a {flat} scan, high,} nearly degenerate masses of {the} dark  particles {are favoured}, {leaving however}  some {viable parameter} space for low dark matter masses ($M_H<\frac{M_h}{2}$).
{The} pair production of dark particles {at the} LHC is {mainly determined}  by their masses {and regions with sizeable cross sections are subject to much more severe limits.} {The current LHC run will hopefully allow to provide more insight into this model, either by strengthening the above constraints or by means of a possible discovery.}

\section*{Acknowledgement}
This work is supported by the 7th Framework Programme of the European Commission through the Initial Training Network HiggsTools PITN-GA-2012-316704. This work was also partly supported by the Polish grant NCN OPUS 2012/05/B/ST2/03306 (2012-2016).
\bibliographystyle{hunsrt}
\bibliography{lit}

\begin{thebibliography}{10}

\bibitem{Ilnicka:2015sra}
Agnieszka Ilnicka, Maria Krawczyk, and Tania Robens.
\newblock {Constraining the Inert Doublet Model}.
\newblock In {\em {2nd Toyama International Workshop on Higgs as a Probe of New
  Physics (HPNP2015) Toyama, Japan, February 11-15, 2015}}, 2015, 1505.04734.

\bibitem{Ilnicka:2015jba}
Agnieszka Ilnicka, Maria Krawczyk, and Tania Robens.
\newblock {The Inert Doublet Model in the light of LHC and astrophysical data
  -- An Update --}.
\newblock 2015, 1508.01671.

\bibitem{Barbieri:2006dq}
Riccardo Barbieri, Lawrence~J. Hall, and Vyacheslav~S. Rychkov.
\newblock {Improved naturalness with a heavy Higgs: An Alternative road to LHC
  physics}.
\newblock {\em Phys.Rev.}, D74:015007, 2006, hep-ph/0603188.

\bibitem{Cao:2007rm}
Qing-Hong Cao, Ernest Ma, and G.~Rajasekaran.
\newblock {Observing the Dark Scalar Doublet and its Impact on the
  Standard-Model Higgs Boson at Colliders}.
\newblock {\em Phys.Rev.}, D76:095011, 2007, 0708.2939.

\bibitem{Dolle:2009ft}
Ethan Dolle, Xinyu Miao, Shufang Su, and Brooks Thomas.
\newblock {Dilepton Signals in the Inert Doublet Model}.
\newblock {\em Phys.Rev.}, D81:035003, 2010, 0909.3094.

\bibitem{Gustafsson:2012aj}
Michael Gustafsson, Sara Rydbeck, Laura Lopez-Honorez, and Erik Lundstrom.
\newblock {Status of the Inert Doublet Model and the Role of multileptons at
  the LHC}.
\newblock {\em Phys.Rev.}, D86:075019, 2012, 1206.6316.

\bibitem{Swiezewska:2012eh}
Bogumila Swiezewska and Maria Krawczyk.
\newblock {Diphoton rate in the inert doublet model with a 125 GeV Higgs
  boson}.
\newblock {\em Phys.Rev.}, D88(3):035019, 2013, 1212.4100.

\bibitem{Arhrib:2013ela}
Abdesslam Arhrib, Yue-Lin~Sming Tsai, Qiang Yuan, and Tzu-Chiang Yuan.
\newblock {An Updated Analysis of Inert Higgs Doublet Model in light of the
  Recent Results from LUX, PLANCK, AMS-02 and LHC}.
\newblock {\em JCAP}, 1406:030, 2014, 1310.0358.

\bibitem{Krawczyk:2013jta}
Maria Krawczyk, Dorota Sokolowska, Pawel Swaczyna, and Bogumila Swiezewska.
\newblock {Constraining Inert Dark Matter by $R_{\gamma\gamma}$ and WMAP data}.
\newblock {\em JHEP}, 1309:055, 2013, 1305.6266.

\bibitem{Chuzhoy:2008zy}
Leonid Chuzhoy and Edward~W. Kolb.
\newblock {Reopening the window on charged dark matter}.
\newblock {\em JCAP}, 0907:014, 2009, 0809.0436.

\bibitem{Swiezewska:2012ej}
Bogumila Swiezewska.
\newblock {Yukawa independent constraints for two-Higgs-doublet models with a
  125 GeV Higgs boson}.
\newblock {\em Phys.Rev.}, D88(5):055027, 2013, 1209.5725.

\bibitem{Gustafsson:2010zz}
Michael Gustafsson.
\newblock {The Inert Doublet Model and Its Phenomenology}.
\newblock {\em PoS}, CHARGED2010:030, 2010, 1106.1719.

\bibitem{Aad:2015zhl}
Georges Aad et~al.
\newblock {Combined Measurement of the Higgs Boson Mass in $pp$ Collisions at
  $\sqrt{s}=7$ and 8 TeV with the ATLAS and CMS Experiments}.
\newblock {\em Phys. Rev. Lett.}, 114:191803, 2015, 1503.07589.

\bibitem{Khachatryan:2014iha}
Vardan Khachatryan et~al.
\newblock {Constraints on the Higgs boson width from off-shell production and
  decay to Z-boson pairs}.
\newblock {\em Phys.Lett.}, B736:64, 2014, 1405.3455.

\bibitem{Aad:2015xua}
Georges Aad et~al.
\newblock {Constraints on the off-shell Higgs boson signal strength in the
  high-mass $ZZ$ and $WW$ final states with the ATLAS detector}.
\newblock {\em Eur. Phys. J.}, C75(7):335, 2015, 1503.01060.

\bibitem{Pierce:2007ut}
Aaron Pierce and Jesse Thaler.
\newblock {Natural Dark Matter from an Unnatural Higgs Boson and New Colored
  Particles at the TeV Scale}.
\newblock {\em JHEP}, 0708:026, 2007, hep-ph/0703056.

\bibitem{Bechtle:2008jh}
Philip Bechtle, Oliver Brein, Sven Heinemeyer, Georg Weiglein, and Karina~E.
  Williams.
\newblock {HiggsBounds: Confronting Arbitrary Higgs Sectors with Exclusion
  Bounds from LEP and the Tevatron}.
\newblock {\em Comput.Phys.Commun.}, 181:138--167, 2010, 0811.4169.

\bibitem{Bechtle:2011sb}
Philip Bechtle, Oliver Brein, Sven Heinemeyer, Georg Weiglein, and Karina~E.
  Williams.
\newblock {HiggsBounds 2.0.0: Confronting Neutral and Charged Higgs Sector
  Predictions with Exclusion Bounds from LEP and the Tevatron}.
\newblock {\em Comput.Phys.Commun.}, 182:2605--2631, 2011, 1102.1898.

\bibitem{Bechtle:2013wla}
Philip Bechtle, Oliver Brein, Sven Heinemeyer, Oscar Stal, Tim Stefaniak,
  et~al.
\newblock {$\mathsf{HiggsBounds}-4$: Improved Tests of Extended Higgs Sectors
  against Exclusion Bounds from LEP, the Tevatron and the LHC}.
\newblock {\em Eur.Phys.J.}, C74(3):2693, 2014, 1311.0055.

\bibitem{Bechtle:2013xfa}
Philip Bechtle, Sven Heinemeyer, Oscar Stal, Tim Stefaniak, and Georg Weiglein.
\newblock {$HiggsSignals$: Confronting arbitrary Higgs sectors with
  measurements at the Tevatron and the LHC}.
\newblock {\em Eur.Phys.J.}, C74(2):2711, 2014, 1305.1933.

\bibitem{Altarelli:1990zd}
Guido Altarelli and Riccardo Barbieri.
\newblock {Vacuum polarization effects of new physics on electroweak
  processes}.
\newblock {\em Phys.~Lett.~B}, 253:161, 1991.

\bibitem{Peskin:1990zt}
Michael~E. Peskin and Tatsu Takeuchi.
\newblock {A New constraint on a strongly interacting Higgs sector}.
\newblock {\em Phys.Rev.Lett.}, 65:964--967, 1990.

\bibitem{Peskin:1991sw}
Michael~E. Peskin and Tatsu Takeuchi.
\newblock {Estimation of oblique electroweak corrections}.
\newblock {\em Phys.Rev.}, D46:381--409, 1992.

\bibitem{Maksymyk:1993zm}
I.~Maksymyk, C.P. Burgess, and David London.
\newblock {Beyond S, T and U}.
\newblock {\em Phys.Rev.}, D50:529--535, 1994, hep-ph/9306267.

\bibitem{Planck:2015xua}
P.A.R. Ade et~al.
\newblock {Planck 2015 results. XIII. Cosmological parameters}.
\newblock 2015, 1502.01589.

\bibitem{Akerib:2013tjd}
D.S. Akerib et~al.
\newblock {First results from the LUX dark matter experiment at the Sanford
  Underground Research Facility}.
\newblock {\em Phys.Rev.Lett.}, 112(9):091303, 2014, 1310.8214.

\bibitem{Lundstrom:2008ai}
Erik Lundstrom, Michael Gustafsson, and Joakim Edsjo.
\newblock {The Inert Doublet Model and LEP II Limits}.
\newblock {\em Phys.Rev.}, D79:035013, 2009, 0810.3924.

\bibitem{Belanger:2015kga}
Genevieve Belanger, Beranger Dumont, Andreas Goudelis, Bjorn Herrmann, Sabine
  Kraml, and Dipan Sengupta.
\newblock {Dilepton constraints in the Inert Doublet Model from Run 1 of the
  LHC}.
\newblock {\em Phys. Rev.}, D91(11):115011, 2015, 1503.07367.

\bibitem{Eriksson:2009ws}
David Eriksson, Johan Rathsman, and Oscar Stal.
\newblock {2HDMC: Two-Higgs-Doublet Model Calculator Physics and Manual}.
\newblock {\em Comput.Phys.Commun.}, 181:189--205, 2010, 0902.0851.

\bibitem{Belanger:2013oya}
G.~Belanger, F.~Boudjema, A.~Pukhov, and A.~Semenov.
\newblock {micrOMEGAs$\_$3: A program for calculating dark matter observables}.
\newblock {\em Comput. Phys. Commun.}, 185:960--985, 2014, 1305.0237.

\bibitem{Alwall:2011uj}
Johan Alwall, Michel Herquet, Fabio Maltoni, Olivier Mattelaer, and Tim
  Stelzer.
\newblock {MadGraph 5 : Going Beyond}.
\newblock {\em JHEP}, 1106:128, 2011, 1106.0522.

\bibitem{Goudelis:2013uca}
A.~Goudelis, B.~Herrmann, and O.~Stal.
\newblock {Dark matter in the Inert Doublet Model after the discovery of a
  Higgs-like boson at the LHC}.
\newblock {\em JHEP}, 1309:106, 2013, 1303.3010.

\bibitem{usbm}
A.~Ilnicka, M.~Krawczyk, and T.~Robens.
\newblock {Inert Doublet Model benchmarks for the 13 TeV run of the LHC}.
\newblock {Submitted to the Higgs Cross Section Working Group, April 2015}.

\end{thebibliography}

\end{document}